\documentclass[preprint,12pt,sort&compress]{elsarticle}

\usepackage{amssymb}
\usepackage{amsmath}
\usepackage{float} 
\usepackage{hyperref}
\usepackage{subfigure}
\usepackage{url}
\usepackage{graphicx}
\usepackage{amsmath}
\usepackage{float}

\usepackage{longtable}
\usepackage{booktabs, tabularx}
\usepackage{adjustbox}
\usepackage{multimedia}
\usepackage[bottom]{footmisc}
\usepackage{footmisc}
\usepackage{graphicx,amsmath,amssymb,bbm,xcolor}
\usepackage{float}

\def\lazz{\mathrel{\mathchoice {\vcenter{\offinterlineskip\halign{\hfil
					$\displaystyle##$\hfil\cr<\cr\sim\cr}}}
		{\vcenter{\offinterlineskip\halign{\hfil$\textstyle##$\hfil\cr<\cr\sim\cr}}}
		{\vcenter{\offinterlineskip\halign{
					\hfil$\scriptstyle##$\hfil\cr<\cr\sim\cr}}}
		{\vcenter{\offinterlineskip\halign{\hfil$\scriptscriptstyle##
					$\hfil\cr<\cr\sim\cr}}}}}
\def\gazz{\mathrel{\mathchoice {\vcenter{\offinterlineskip\halign{\hfil
					$\displaystyle##$\hfil\cr>\cr\sim\cr}}}
		{\vcenter{\offinterlineskip\halign{\hfil$\textstyle##$\hfil\cr>\cr\sim\cr}}}
		{\vcenter{\offinterlineskip\halign{
					\hfil$\scriptstyle##$\hfil\cr>\cr\sim\cr}}}
		{\vcenter{\offinterlineskip\halign{\hfil$\scriptscriptstyle##
					$\hfil\cr>\cr\sim\cr}}}}}		
\def\pr{\prime}
\def\be{\begin{equation}}

\def\ran{\right\rangle}
\def\ee{\end{equation}}
\def\barr{\begin{array}}
	\def\earr{\end{array}}

\def\l{\left}
\def\r{\right}
\def\dis{\displaystyle}
\def\ed{\end{document}}

\def\cs{{\bf s}}

\begin{document}

\begin{frontmatter}

\title{Distribution of lowest eigenvalue in $k$-body bosonic random matrix ensembles}

\author[add1]{N. D. Chavda}
\author[add1]{Priyanka Rao\corref{cor1}}
\author[add2]{V. K. B. Kota}
\author[add3]{Manan Vyas\corref{cor2}}

\address[add1]{Department of Applied Physics, Faculty of Technology and Engineering, 
The Maharaja Sayajirao University of Baroda, Vadodara-390001, India}
\address[add2]{Physical Research Laboratory,  Ahmedabad 380 009, India}
\address[add3]{Instituto de Ciencias F{\'i}sicas, Universidad Nacional Aut{\'o}noma de M\'{e}xico, 62210 Cuernavaca, M\'{e}xico}

\cortext[cor1]{Present address: Department of Statistical Science, Duke University, Durham, NC 27708-0251, USA}
\cortext[cor2]{corresponding author, manan@icf.unam.mx}

\begin{abstract}

We present numerical investigations demonstrating the result that the distribution of the lowest eigenvalue of 
finite many-boson systems (say we have $m$ number of bosons) with $k$-body interactions, modeled by Bosonic 
Embedded Gaussian Orthogonal [BEGOE($k$)] and Unitary [BEGUE($k$)] random matrix Ensembles of $k$-body 
interactions, exhibits a smooth transition from Gaussian like (for $k=1$) to a modified Gumbel like (for intermediate 
values of $k$) to the well-known Tracy-Widom distribution (for $k=m$) form. We also provide ansatz for centroids 
and variances of the lowest eigenvalue distributions. In addition, we show that the distribution of normalized spacing between the lowest and next lowest eigenvalues exhibits a transition from Wigner’s surmise (for $k = 1$) to Poisson (for intermediate $k$ values with $k \le m/2$) to Wigner’s surmise (starting from $k = m/2$ to 
$k = m$) form.  We analyze these transitions as a function of $q$ parameter defining $q$-normal distribution for eigenvalue densities.

\end{abstract}


\end{frontmatter}

\section{Introduction} \label{intro}

Extreme value statistics (EVS) is related to, for example,  the statistics of either the lowest few or largest few eigenvalues of a matrix and has found varied applications \cite{Bo-97, De-08, Fo-15}.  Depending on the parent distribution, the classical EVS are classified into Fréchet, Gumbel and Weibull distributions \cite{Gumbel-1958, Galambos-1987}.  The classical EVS deals with the statistics of minimum or maximum of a set of independent random variables with a given parent distribution whereas in most of the real physical systems, the underlying variables are correlated \cite{Ma-20}.  Complex systems are usually modeled by classical random matrix ensembles - Gaussian Orthogonal (GOE), Unitary (GUE) and symplectic (GSE) Ensembles \cite{Mehta, RMT-book}.  The eigenvalues are correlated for these ensembles and the EVS of the (lowest) largest eigenvalues is described by the celebrated (reflected) Tracy-Widom (TW) distribution \cite{TW-1, TW-2, TW-3}.  For a variety of random matrix ensembles, EVS has been investigated; see for example \cite{Ch-14, Sa-24}.  However, one class of ensembles where there has been very little attention are embedded random matrix ensembles with $k$-body interactions which are now established to be essential while dealing with dynamics of complex quantum many-body systems \cite{MF,  Br-81, BW, vkbk,Ma-2017}.  

Embedded ensembles (EE) are random matrix ensembles that describe the generic properties of complex interacting many-particle (fermion/boson) systems \cite{vkbk}.  Given $m$ number of fermions or bosons distributed in $N$ single particle levels interacting via $k$-body interactions ($k \leq m$),  the $k$-body fermionic or bosonic EE are constructed by defining the $k$-particle Hamiltonian to be a GOE (or a GUE) and then propagating it to $m$-particle spaces using the underlying Lie algebra \cite{vkbk}.  The case when rank of interactions $k$ equals number of fermions or bosons $m$, we have a GOE (or a GUE).  EVS is well-understood in this case as the matrix elements are independent and identically distributed (i.i.d.) random variables \cite{BB-91, Fo-93, TW-1, TW-2, TW-3}.  As the $k$-particle Hamiltonian is embedded in the $m$-particle Hamiltonian, the many-particle matrix elements are correlated for EE, unlike a GOE (or a GUE).  Note that although the matrix elements for GOE (or a GUE) are i.i.d.  random variables, its eigenvalues are correlated.  Therefore,  the eigenvalues of EE will have additional (complex) correlations and we expect deviations from the TW distribution for EE with $k < m$.  There are a few examples where one can derive EVS in presence of correlations between matrix elements \cite{PRL-23, Sa-24}.

EE are paradigmatic random matrix models which incorporate the few-particle interaction dominance in many-particle systems and hence, the Fock-space sparsity of quantum many-body systems \cite{MF,  Br-81, BW, vkbk,Ma-2017, Ko-01,  Be-01, Al-00, Zh-04, Pa-07, Gu-98, Ko-11, Sm-15, Zel-16, Pa-21,PRE-22}.  They provide a more accurate description of generic spectral properties of many-particle interacting quantum systems in the chaotic regime.  See also Refs.  \cite{MF,  Br-81, BW, vkbk, Ma-2017, Ko-01, Be-01, Al-00, Zh-04, Pa-07, Gu-98, Ko-11, Sm-15, Zel-16, Pa-21, PRE-22} for details regarding various properties and applications of EE.  In addition,  Sachdev–Ye–Kitaev (SYK) models that have been receiving increasing attention in high-energy physics in recent years \cite{Ki-15a, Ki-15b, Sa-93, Gar-16, Gar-17,Gar-19, Gu-20, Jia-20, Al-24, Gar-24} are also examples of EE with complex fermions replaced by Majorana fermions.  Moreover, EE are generic, though analytically difficult to deal with, compared to for example lattice spin models \cite{Ma-14,  Pa-21} as the latter are associated with spatial coordinates (nearest and next-nearest neighbor interactions) only.  Also,  the universal properties derived using EE extend easily to systems represented by lattice spin models, for example eigenvalue density is Gaussian in the mean-field basis unlike a classical GOE \cite{Ma-14, Pa-21}. It will be interesting to explore possibility of realization of EE using quantum computers \cite{QC-24}.

In addition,  it is relevant and important to analyze how the distribution of the largest or smallest eigenvalue for $k$-body fermionic or bosonic EE varies with $k$, the rank of the interactions, for a $m$ particle (fermion or boson) system as $k$ changes from 1 to $m$. In a first study \cite{IJMPE-2018},  for {\it two-body} fermionic and bosonic EE,  the distribution of the lowest eigenvalue is studied numerically and it is shown to follow the modified-Gumbel distribution that was used earlier in \cite{Pa-08}.  Details of modified-Gumbel distribution will be given in Section \ref{sec1} ahead.  Going beyond two-body ensembles, recently first numerical results for the distribution of largest eigenvalue (same results expected for the lowest eigenvalue) for Fermionic Embedded Gaussian Orthogonal [FEGOE($k$)] and Unitary [FEGUE($k$)] random matrix Ensembles with $k$-body interactions are presented in \cite{Be-23}.  It is shown that the distribution exhibits a smooth transition between Gaussian and TW form as $k$ changes from 1 to $m$ for the $m$ fermion systems considered.  However,  the scaling employed for the eigenvalues depends only on the system dimensions and thus, does not depend on the rank of interactions $k$.   In the present paper, we focus on Bosonic Embedded Gaussian Orthogonal [BEGOE($k$)] and Unitary [BEGUE($k$)] random matrix Ensembles with $k$-body interactions.  BEGOE($k$)/BEGUE($k$) are closely connected to interacting boson models of atomic nuclei \cite{Ma-12n},  interacting boson models like Bose-Hubbard models \cite{Pa-21},  symmetrized two-coupled rotors model \cite{Bor-98},  {\it etc}.  Our aim is not only to present results for EVS for bosonic EE with $k$-body interactions for a $m$ particle system (with $k$ changing from 1 to $m$) but also, more importantly,  it is to bring out the role of the so-called $q$ parameter that is central to $k$-body EE ($q$ changes from 1 to 0 as $k$ changes from 1 to $m$).  Following the result from SYK model due to Verbaarschot and collaborators \cite{Gar-17}, the role of parameter $q$ (or $q$-normal distribution) in EE was investigated in a series of papers \cite{JSM-2019, Rao-21,  JSM-21, Rao-23, Ko-23}.  We will briefly turn to this now.

Significantly, the transition in eigenvalue density, as $k$ changes from 1 to $m$, for BEGOE($k$) and BEGUE($k$) [also for FEGOE($k$) and FEGUE($k$)] is well described by the $q$-normal form \cite{Ismail, Sza-1} with the parameter $q$ being related to the fourth moment of the eigenvalue density as shown in \cite{JSM-2019}.  Note that, for $q = 1$ the $q$-normal reduces to Gaussian eigenvalue density and this is known to be valid for $k << m/2$ \cite{MF} and similarly,  for $q = 0$ the $q$-normal gives the well known semi-circle (GOE/GUE) eigenvalue density as valid for $k = m$.  Going beyond this important result in the study of EE, the parameter $q$ is also shown to play an important role in determining many other properties of $k$-body EE \cite{JSM-2019, Rao-21,  JSM-21, Rao-23, Ko-23,  Man-23}.  Our interest here is to study the distribution of smallest (or, equivalently largest) eigenvalues for EE($k$). However, as already realized in \cite{IJMPE-2018, Be-23},  at present no analytic approach that gives this distribution is available either for fermionic and bosonic EE as the matrix elements in these ensembles are correlated  \cite{BW, vkbk, PRL-23, Sa-24}.  For the BEGOE($k$) and BEGUE($k$) considered in this paper,  we provide ansatz for centroids and variances of these distributions and they match very well with the numerical results obtained.  However,  as further analytics are difficult at this point of time, we analyze these distributions numerically.  We would like to point out that though we choose some smaller systems (small values of $N$ and $m$) to compare the numerically obtained moments with the proposed ansatz,  we have also performed numerically extensive calculations with $N = 5$ and $m=10$ for 5000 member ensembles for obtaining a plausible result for the lowest eigenvalue distribution.  In particular, we have numerically investigated how the distribution of the lowest eigenvalues for BEGOE($k$) and BEGUE($k$) varies as a function of the $q$ parameter.  For $k = m$ (GOE/GUE), the distribution of the lowest eigenvalues converges to the well-known TW distribution \cite{TW-1, TW-2, TW-3}.  Going beyond the results presented for fermionic EE in \cite{Be-23},  we provide further analytical insight by connecting our numerical results to the $q$-normal distribution for the eigenvalue density and discuss not only the limiting distributions for non-interacting ($k = 1$) and maximal interaction order ($k = m$) but also the entire transition in between, where we also identify the underlying probability distribution.

Now, we will give a preview.  Section \ref{sec1} defines the BEGOE($k$)/BEGUE($k$) and gives the $q$-normal form for the eigenvalue densities along with the formula for the parameter $q$ in terms of $(N,m,k)$.  We also give the various distributions that have been used in the present work to analyze the distribution of lowest eigenvalues.  In Section \ref{sec2}, we then analyze the first four moments of the lowest eigenvalue distribution.  We compare the lowest eigenvalue distribution with EVS and study the spacing distribution between lowest eigenvalue and its nearest neighbor in Section \ref{sec3}.  Finally, Section \ref{sec4} gives conclusions and future outlook.

\section{Preliminaries}
\label{sec1}

In this section, we define bosonic EE and give the $q$-normal form for the eigenvalue densities \cite{JSM-2019}. Then we explain the various distributions that have been used to analyze the distribution of  lowest eigenvalues for BEGOE($k$) and BEGUE($k$). 

\subsection{BEGOE($k$) and BEGUE($k$) ensembles}

Given a system of $m$ spin-less bosons in $N$ degenerate single particle states and say the interaction among the bosons is a $k$-body ($1 \leq k \leq m$) in nature,  then the Hamiltonian operator for the system takes the form 
\be
H(k,\beta) = \dis\sum_{k_a,k_b} v_{k_a,k_b}^{\beta} B^\dagger(k_a) B(k_b)\;.
\label{eq.ent1}
\ee
Here, $k_a$ and $k_b$ denote $k$-particle configuration states in occupation number basis and 
$B^\dagger(k_a)$ creates a normalized $k$ particle state $\l.\l|k_a\r.\ran$ with $B^\dagger(k_a)\l.\l|0\r.\ran=\l.\l|k_a\r.\ran$. Similarly $B(k_b)$ is a $k$ particle annihilation operator.  Note that $v_{k_a,k_b}^{\beta}$ are the matrix elements of $H$ in the defining $k$ particle space with the $H$ matrix dimension being $d_k={N+k-1 \choose k}$.  Here,  Dyson's parameter $\beta$ is equal to $1$ for GOE and $2$ for GUE.  Now, representing the $v$ matrix by GOE/GUE in $k$ particle space we have a GOE/GUE ensemble of operators and action of each member of this GOE/GUE on the $m$ particle states will generate a $m$ particle matrix of dimension $d_m={N+m-1 \choose m}$. The ensemble of these matrices form embedded GOE/GUE of $k$ particle interactions [BEGOE($k$)/BEGUE($k$) with $B$ for bosons] in $m$ particle spaces.  In defining the GOE in $k$-particle spaces, we choose the matrix elements to be independent Gaussian random variables with variance 2 for diagonal matrix elements and 1 for off-diagonal matrix elements. For GUE, the variance of real and imaginary parts of the off-diagonal matrix elements are chosen to be unity.  For the BEGOE($k$) and BEGUE($k$), the ensemble averaged odd moments of the eigenvalue density are all zero by definition. The lowest scaled even moments of interest are the fourth ($\mu_4$), sixth ($\mu_6$) and eight ($\mu_8$) moment. Using the formulas for these moments derived in \cite{MF,  Sm-15},  it is conclusively established in \cite{JSM-2019} that the eigenvalue density for both these ensemble takes $q$-normal form.  Briefly,  the $q$-normal form is as follows.

\subsection{$q$-normal form for eigenvalue density}

In order to introduce the $q$-normal form, firstly one needs the definition of $q$ numbers $[n]_q$ and they are(with $[0]_q=0$), 
\be
\l[n\r]_q = \dis\frac{1-q^n}{1-q} = 1+q + q^2 + \ldots+q^{n-1}\;.
\label{eq.q1}
\ee
Note that $[n]_{q \rightarrow1}=n$. Similarly, $q$-factorial $[n]_q! = \dis\Pi^{n}_{j=1} \,[j]_q$ with $[0]_q!=1$. Given these, the $q$-normal distribution $f_{qN}(x|q)$ with $x$ being a standardized variable (then $x$ is zero centered with variance unity), is given by \cite{Ismail,Sza-1} 
\be
f_{qN}(x|q) = \dis\frac{\dis\sqrt{1-q} \dis\prod_{k^\pr=0}^{\infty} \l(1-
q^{k^\pr +1}\r)}{2\pi\,\dis\sqrt{4-(1-q)x^2}}\; \dis\prod_{k^\pr=0}^{\infty}
\l[(1+q^{k^\pr})^2 - (1-q) q^{k^\pr} x^2\r]\;.
\label{eq.q2}
\ee
The $f_{qN}(x|q)$ is non-zero for $x$ in the domain defined by $\cs(q)$ where
\be
\cs(q) = \l(-\dis\frac{2}{\dis\sqrt{1-q}}\;,\;+\dis\frac{2}{\dis\sqrt{1-q}}\r)\;.
\label{eq.q3}
\ee
Note that $\int_{\cs(q)} f_{qN}(x|q)\,dx =1$. Most important property of $q$-normal is that $f_{qN}(x|1)$ is Gaussian with $\cs(q=1)=(-\infty , \infty)$ and similarly, $f_{qN}(x|0)=(1/2\pi) \sqrt{4-x^2}$, the semi-circle with $\cs(q=0)=(-2,2)$. 

It is also useful to mention that $f_{qN}(x|q)$ is the weight function with respect to which the $q$-Hermite polynomials are orthogonal over $S(q)$ giving,
$$
\dis\int_{s(q)} H_n(x|q) \; H_m(x|q) \; f_{qN}(x|q) \; dx =[n]_{q}! \, \delta_{mn} \;,
$$
and the $q$-Hermite polynomials satisfy the 3-term recursion relation,
$$
H_{n+1}(x|q)=x\,H_n(x|q) - [n]_q\,H_{n-1}(x|q)\;\;\mbox{with}\;\;
n \geq 1,\; H_{-1}(x|q)=0,\;H_0(x|q)=1 \;.
$$

The reduced central fourth moment for eigenvalue density is $\gamma_2(N,m,k) = q(N,m,k) - 1$ and the formula for parameter $q(N,m,k)$ for BEGOE($k$)/BEGUE($k$) is \cite{JSM-2019},
\be
\barr{l}
q(N,m,k) = \dis\binom{N+m-1}{m}^{-1} \dis\sum_{\nu=0}^{\nu_{max}}\; \dis\frac{
\Lambda_B^\nu(N,m,m-k)\;\Lambda_B^\nu(N,m,k) \;d_B(g_\nu)}{\l[\Lambda_B^0(N,m,k)\r]^2}
\\ \\
\Lambda_B^\nu(N,m,r) =  \dis\binom{m-\nu}{r}\;\dis\binom{N+m+\nu-1}{r}\;,\\ \\
d_B(g_\nu)  = \dis\binom{N+\nu-1}{\nu}^2-\dis\binom{N+\nu-2}{\nu-1}^2\;.
\earr \label{eq.qh9b}
\ee
Here, $\nu_{max} = min(k,m-k)$. This formula is pretty accurate for $k \geq 2$ but there are large deviations for $k =1$  \cite{JSM-2019, Rao-21}.  Note that $\Lambda_B^0(N,m,k)$ gives the spectral variance.

\subsection{Extreme value statistics}

In the present work,  following the results in \cite{TW-1, IJMPE-2018, Pa-08, Be-23}, we use three different distributions that characterize EVS.  These are: Gaussian (${\cal G}$),  Classical TW and Modified Gumbel distribution.  Out of the three distributions we have used,  Gaussian is simplest and for lowest eigenvalue $E$ with zero center and unit variance we have ${\cal G}(E) = (2\pi)^{-1/2}\;\exp (-E^2/2)$. 

\subsubsection{Classical Tracy-Widom distribution}

In the large $D$ limit,  the TW distribution of lowest eigenvalues $E_{min}$, corresponding to a $D$-dimensional system,  is given by $F_\beta(\widehat{E})$ \cite{TW-1, TW-2, TW-3},  with the normalized eigenvalues $\widehat{E}$ defined as
\begin{equation}
\widehat{E} = D^{1/6} (E_{min} + 2\sqrt{\beta D}) \;.
\end{equation}
For $\beta = 1$ and 2, we have
\begin{equation}
\begin{array}{l}
F_1(x) = \exp[\Theta(x)] \; F_2^{1/2}(x) \;, 
\\ \\
F_2(x) = \exp \left( - \displaystyle\int_x^\infty (s-x) [Q(s)]^2 \; ds\right) \;.
\end{array}
\label{eq-tw}
\end{equation}
Here, $Q$ is given in terms of the solution to Painlevé type II equation $Q^{\prime\prime} = sQ + 2Q^3$ subjected to the boundary condition $Q(s) \approx Ai(s)$ for $s \to \infty$, with $Ai(s)$ denoting the Airy function and $\Theta(x) = -\int_x^\infty Q(s) \; ds/2$.  Note that $\beta$ is the Dyson's parameter.

\subsubsection{Modified Gumbel distribution}

Gumbel distributions are one of the EVS and have been used to analyze ground state distribution in Sherrington-Kirkpatrick model  \cite{Pa-08} and Two-body random ensembles (TBRE) \cite{IJMPE-2018}.  The modified Gumbel distribution is given by \cite{Gumbel-1958, Galambos-1987},
\be
G_{\mu}(E)= w \exp \l[\mu\left(\frac{E-u}{v}\right)-\mu\exp \left(
\frac{E-u}{v}\right)\r] \;.
\label{gumbel}
\ee
Note that $u$ and $v$ are rescaling parameters and $w$ is a normalization constant.  The ensemble averaged data are compared with Eq. \eqref{gumbel} by shifting and scaling the distribution such that the mean is zero and the variance is unity which fixes $u$, $v$ and $w$ as a functions of parameter $\mu$.  The $G_{\mu}$ provides an interpolation between the standard Gumbel ($\mu=1$) and Gaussian ($\mu =\infty$) distribution.  It is conjectured in \cite{Bram-00} that the $G_{\mu}$ distribution with $\mu=\pi/2$ applies to many correlated  systems.  Also, the $G_{\mu}$ distribution with non-integer $\mu$ value was used in the study of three dimensional spin glasses \cite{Berg-02}.  All these give a good reason for using $G_{\mu}$ distribution in the present analysis.  

\section{First four moments of the lowest eigenvalue distribution}
\label{sec2}

In all the examples considered in this paper, we construct a 1000 member BEGOE($k$) and BEGUE($k$) with following choices of parameters $N$ and $m$:  $N=4$, $m=4-14$; $N=5$, $m=5-11$; and $N=6$, $m=6-9$.  Remember that rank of interactions $k$ takes values from 1 to $m$.  In all these examples, we obtain the lowest eigenvalue; we denote this by $\lambda$ hereafter.  For the distribution of $\lambda$, we will present in this section,  results for the centroid $\lambda_c$, width $\sigma_\lambda$, skewness 
$S$ (defined by the third central moment) and kurtosis $\kappa$ (defined by the fourth central moment).

\begin{figure}[tbh]
	\begin{center}
		\includegraphics[width=0.65\linewidth]{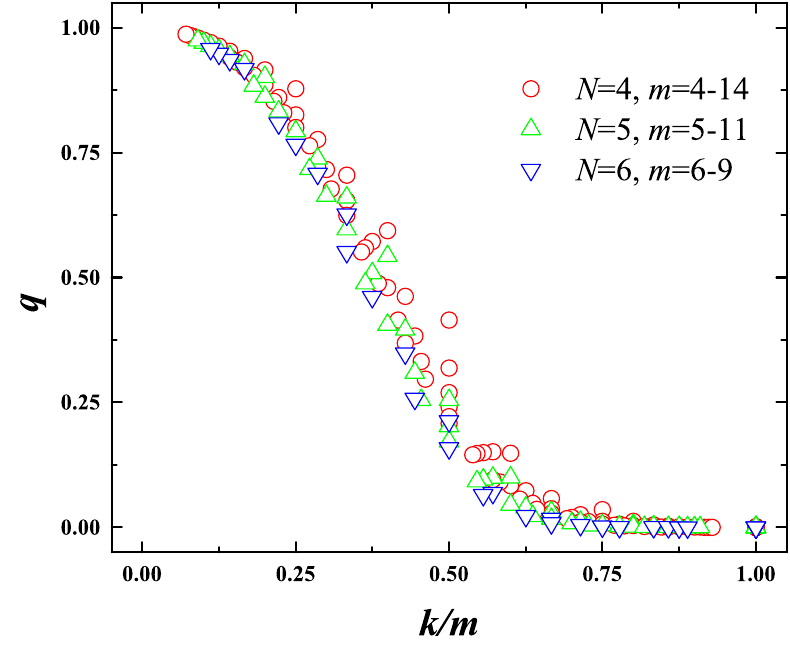}
	\end{center}
	\caption{Variation of parameter $q(N,m,k)$ as a function of $k/m$ for various values of $(N,m): $ $N=4$, $m=4-14$; $N=5$, $m=5-11$; and $N=6$, $m=6-9$.}
	\label{fig1}
\end{figure}

Firstly, we compute the parameter $q$ using Eq. \eqref{eq.qh9b} and its variation with $k$ is shown in Fig. \ref{fig1}. Notice the linear behavior for intermediate values of $k/m$ while there are strong deviations near the two extreme values of $k$. Although the variation of $q$ with $k/m$ is smooth,  there is very weak dependence on $N$. 

Qualitatively,  larger(smaller) values of the parameter $q$ implies pre-dominance of few-body(many-body) interactions in the complex system under consideration.  In the past studies \cite{JSM-2019, Rao-21,  JSM-21, Rao-23, Ko-23,  Man-23}, it has been shown that the $q$ parameter explains bulk properties of the spectra very well in a unified way.  In the present paper, we want to analyze its role in understanding the distribution of minimum eigenvalues (far away from the bulk) and present approximate results for its moments. 

\subsection{Centroid}

\begin{figure}[tbh]
	\begin{center}
		\begin{tabular}{cc}
			\includegraphics[width=0.45\linewidth]{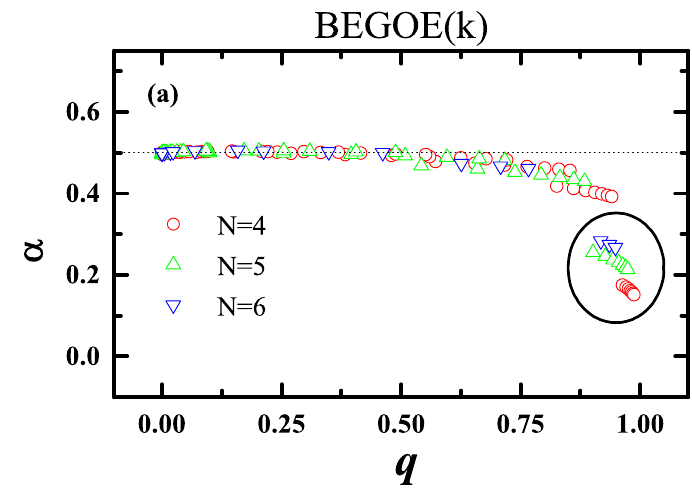}&\includegraphics[width=0.45\linewidth]{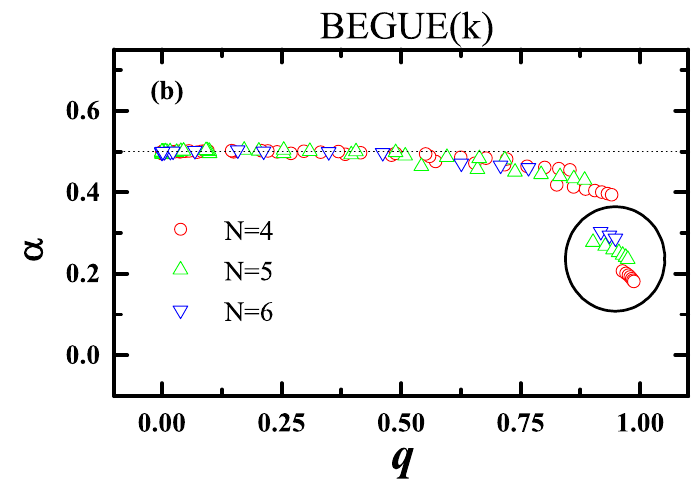}
		\end{tabular}
	\end{center}
	\caption{Variation in parameter $\alpha$ defined by ansatz given in Eq. \eqref{eq-alp} for the centroids of the lowest eigenvalue distributions for a 1000 member (a) BEGOE($k$) and (b) BEGUE($k$) with different values of $(N,m)$ ($m$ values corresponding to each $N$ are given in Fig. \ref{fig1}) and $k = 1, \;2,\;3,\ldots,\;m$.  The ovals highlight the points corresponding to $k = 1$. }
	\label{fig2}
\end{figure}

The $q$-normal distribution shows that the distribution has a cut-off at $-2 [\beta \Lambda_B^{0}(N,m,k)]^{1/2}/[1-q(N,m,k)]^{1/2}$ at the lower edge when measured with respect to the ensemble averaged eigenvalue centroid; this follows easily from Eq. \eqref{eq.q3}.  For finite $(N,m)$, there will be departures as $q$-normal is an asymptotic form for the eigenvalue densities. Therefore, we use the following parametrization for the centroid $\lambda_c$ of $\lambda$'s for a given $(N,m,k)$,
\begin{equation}
\lambda_c(N,m,k)  = \displaystyle\frac{-2}{\displaystyle\sqrt{1- q(N,m,k)}}{\left[ \beta \Lambda_B^{0}(N,m,k) \right]}^{\alpha}.
\label{eq-alp}
\end{equation}
Note that for both BEGOE($k$) and BEGUE($k$), formulas for $q(N,m,k)$ and for $\Lambda^0_B(N,m,k)$ follow from Eq. \eqref{eq.qh9b}.  In the limit $q$-normal form is exact, the parameter $\alpha = 1/2$ and secondly, for $k = m$,  by applying Eq. \eqref{eq.qh9b}, Eq. \eqref{eq-alp} gives the well-known result for TW for GOE/GUE. Therefore, using the calculated $\lambda$ values, via least-square procedure, we have determined the values of $\alpha$ for all $(N,m,k)$ values listed above.  The corresponding results are shown in Fig. \ref{fig2}. For $0 \leq q \leq 0.75$, the agreement with $q$-normal form result that $\alpha = 0.5$ is almost exact. However,  for $q(N,m,k) \gazz 0.8$, the deviations from $\alpha = 1/2$ are significant and they correspond to $k = 1$.  This appears to be due to the well known result that the eigenvalue centroid scaled by the spectral width fluctuations from member to member are largest for $k =1$ \cite{Chavda2003} and they decrease faster as $k$ increases. This also corresponds to ensemble vs spectral averaging in EE \cite{Brody, SeligmanFlores}. This effect of non-ergodicity is discussed in Sec. \ref{app1} below.   In addition, the $N = 4$ systems show much larger deviations from $\alpha = 0.5$. This is understandable as one needs at least five single particle states for asymptotics to work well \cite{KPat-2000-PLA}.  

\subsubsection{Fluctuations in $\alpha$ values for centroids: Effect of non-ergodicity}
\label{app1}

\begin{figure}[H]
	\begin{center}
		\includegraphics[width=0.65\linewidth]{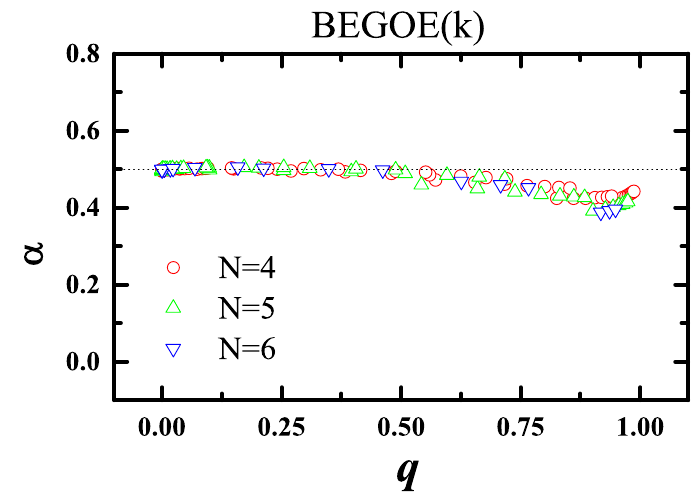}
	\end{center}
	\caption{Variation in parameter $\alpha$ for the centroids $\lambda_c(N,m,k)$ of the lowest eigenvalue distributions for a 1000 member BEGOE($k$) with different values of $(N,m)$ ($m$ values corresponding to each $N$ are given in Fig. \ref{fig1}) and $k = 1, 2, \ldots,m$.  We have computed the parameter $q_i(N, m,k)$ for each member $i$ of the ensemble using the kurtosis,  accounting for the member-to-member fluctuations (non-ergodicity) in lowest eigenvalues $\lambda_i(N,m,k)$ and $q_i(N,m,k)$ parameter,  unlike in Fig. \ref{fig2}.}
		\label{fig9}
\end{figure}

To further probe into the fluctuations in $\alpha$ values from $0.5$ as shown in Fig. \ref{fig2}, we computed the parameter $q_i(N, m,k)$ for each member $i$ of the ensemble using the kurtosis. Then, used $q_i(N, m,k)$ along with the minimum eigenvalue $\lambda_i(N,m,k)$ for each member to calculate $\lambda_c(N,m,k) = \langle \lambda_i(N,m,k) \sqrt{1-q_i(N,m,k)} \rangle$.  Then the exponent $\alpha$ is obtained using the equation $\lambda_c(N,m,k) = -2.0 {[\beta \Lambda^{0}(N,m,k)]}^{\alpha}$.  The results for a 1000 member BEGOE($k$) are shown in Fig. \ref{fig9}.  One can see that $\alpha \sim 0.5$ for all $k$ values, unlike as in Fig. \ref{fig2}. Thus, the member-to-member fluctuations (non-ergodicity) in the minimum eigenvalues $\lambda_i(N,m,k)$  and parameter $q_i(N,m,k)$ are giving the large deviations for $\alpha$ from $0.5$ in Fig. \ref{fig2}.  Though not shown,  similar results are obtained also for BEGUE($k$).  The systems with $k=1$ appear to be special and there are results in \cite{Majumdar-NIP} for different types of $k=1$ interactions. 

\subsection{Variance}

The variance for TW distribution is $D^{-1/6}$ for $D$-dimensional GOE/GUE. This corresponds to BEGOE($k$)/BEGUE($k$) with $k = m$ and $D = \binom{N+k-1}{k}$. This is also proportional to the spectral width. Therefore, we use the following two parameterizations for bosonic EE, 
\be
\sigma_{\lambda} (N,m,k) = \left[\Lambda^{0}_B(N,m,k) \right]^{\mu_1},
\label{eq-mu1}
\ee
and
\be
\sigma_{\lambda}(N,m,k) = {\left[\Lambda^{0}_B(N,m,k) \right]}^{\mu_2} {N+k-1 \choose k}^{(-1/2)}.
\label{eq-mu2}
\ee
For $k=m$ case, i.e. for classical Gaussian ensembles, the exponent $\mu_1=-1/6$ for Eq. \eqref{eq-mu1} and $\mu_2=1/3$ for Eq. \eqref{eq-mu2}. Both of these will give $\sigma_{\lambda}=D^{-1/6}$ for the case $k=m$.

\begin{figure}[H]
	\begin{center}
		\begin{tabular}{cc}
			\includegraphics[width=0.45\linewidth]{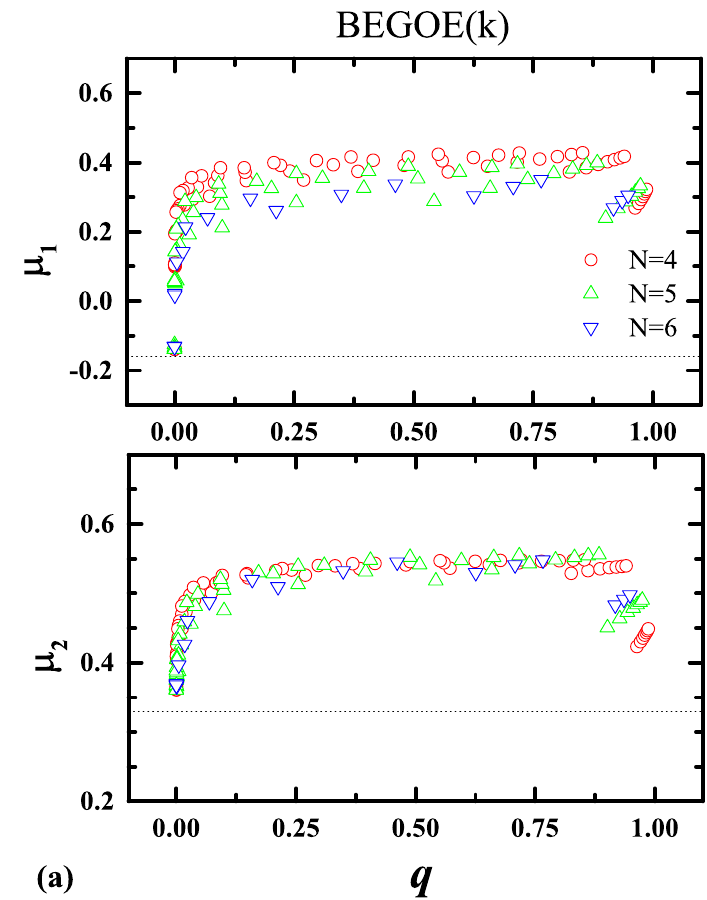}&
                          \includegraphics[width=0.45\linewidth]{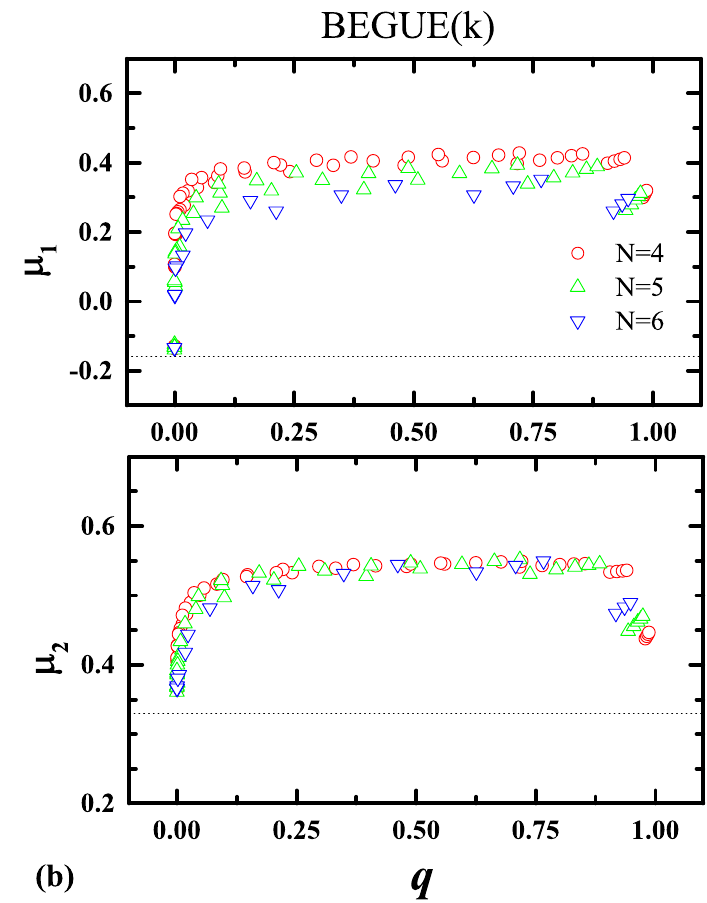}
		\end{tabular}
	\end{center}
	\caption{Variation in the parameters $\mu_1$ and $\mu_2$ defined by ansatz given respectively in Eqs. \eqref{eq-mu1} and \eqref{eq-mu2} for the variances of the lowest eigenvalue distributions for a 1000 member  (a) BEGOE($k$) and (b) BEGUE($k$) as a function of parameter $q(N,m,k)$,  with different values of $(N,m)$ ($m$ values corresponding to each $N$ are given in Fig. \ref{fig1}) and $k = 1, \;2,\ldots,\;m$.  }
		\label{fig3}	
\end{figure}

Now, using the calculated $\lambda$ values, via least-square procedure, we have determined the values of $\mu_1$ and $\mu_2$ for all $(N,m,k)$ values listed above. The results are shown in Fig. \ref{fig3}. For small values of $q \lazz 0.1$,  there is a sharp increase in the values of $\mu_1$ and $\mu_2$ and then,  essentially become a constant.  The values of $\mu_1$ and $\mu_2$ decrease for $k =1$.  Note that $\mu_1$ becomes positive (also $\mu_2$ increases) which implies larger variance compared to that for TW. This trend is common for both EGOE/EGUE \cite{Be-23}.

\subsection{Skewness and excess}

\begin{figure}[H]
	\begin{center}
			\includegraphics[width=0.45\linewidth]{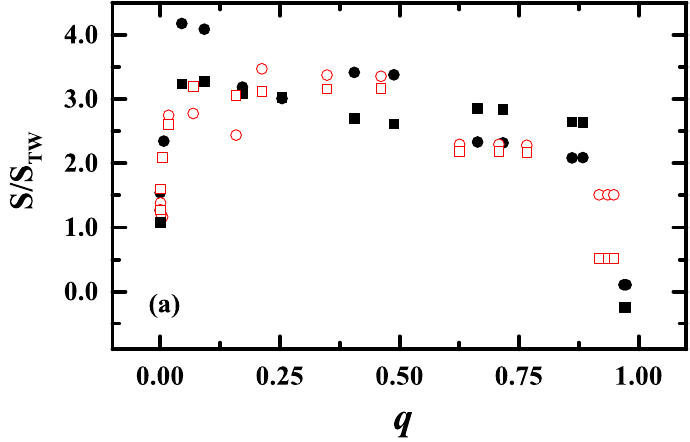}
			\includegraphics[width=0.45\linewidth]{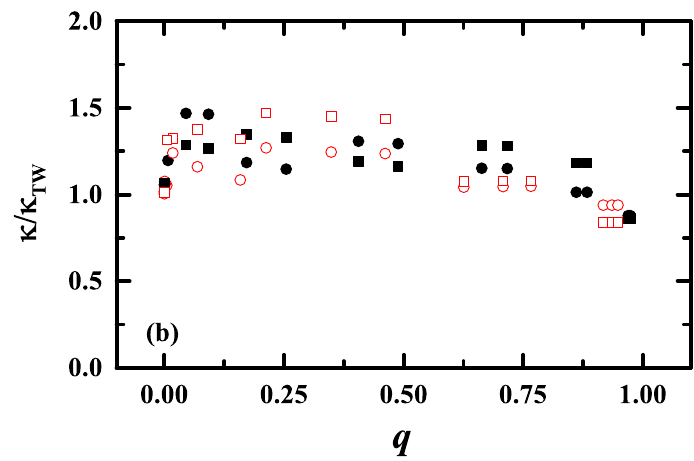}
	\end{center}
	\caption{Variation in skewness (left panel) and kurtosis (right panel) for BEGOE($k$) (squares) and BEGUE($k$) (circles) as a function of parameter $q(N,m,k)$.  Results are shown for 1000 member ensembles with $N=5$, $m=10-11$ (solid symbols) and $N=6$, $m=6-8$ (open symbols).  Note that $S_{TW}=-0.2935$, $\kappa_{TW}=3.1652$ for $\beta=1$ and $S_{TW}=-0.2241$,  $\kappa_{TW}=3.0934$ for $\beta=2$.}
		\label{fig4}
\end{figure}

We have computed the shape parameters - skewness $S$ and kurtosis $\kappa$ for the $\lambda$ distribution and analyze their variation with parameter $q$ in Fig. \ref{fig4} using some examples. For very small and very large $q$ values, we see departure from the linear decreasing behavior in both skewness and kurtosis. For $0.1 \lazz q \lazz 0.8$, the $S/S_{TW}$ value decreases from $\sim$ 4 to 2 while $\kappa/\kappa_{TW}$ value decreases from $\sim$ 1.5 to 1.0. Thus, the skewness and kurtosis values are larger than those for TW distribution for intermediate $q$ values (see $k = m$ result in Fig. \ref{fig4}).

It might be insightful to derive the expressions for variance, skewness and kurtosis from first principles but this is beyond the scope of the present paper.

\section{Lowest eigenvalue distribution: Comparison with EVS}
\label{sec3}

\begin{figure}[tbh]
		\includegraphics[width=\linewidth]{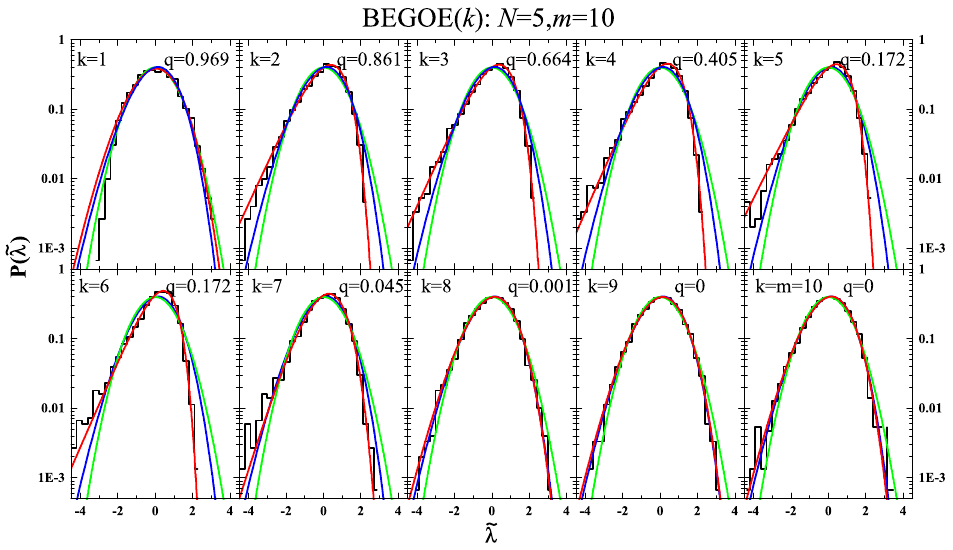}
	\caption{Normalized probability distributions $P({\tilde\lambda})$ for the lowest eigenvalues for a 5000 member BEGOE($k$) with $(N,m) = (5,10)$. The Gaussian (smooth green curves), modified Gumbel (smooth red curves) distribution using Eq. \eqref{gumbel} and the TW (smooth blue curves) distribution using Eq. \eqref{eq-tw} are superimposed in each panel with the numerical histograms.}
	\label{fig5}
\end{figure}

\begin{figure}[tbh]
		\includegraphics[width=\linewidth]{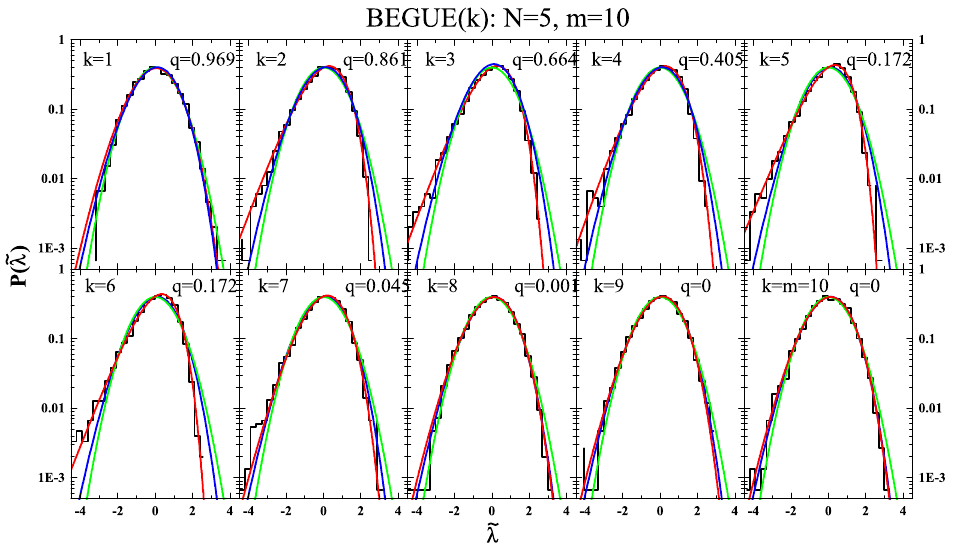}
	\caption{Same as Fig.~\ref{fig5} but for BEGUE($k$).}
	\label{fig6}
\end{figure}

In this section, we compare the distribution of lowest eigenvalues with EVS given in Section \ref{sec1}. The results for a 5000 member BEGOE($k$) and BEGUE($k$) are given in Figs. \ref{fig5} and \ref{fig6}, respectively. These are obtained for $N = 5$ and $m = 10$ system with $k$ varying from 1 to 10. The numerical histograms are computed using the scaled lowest eigenvalues 
\be
\tilde{\lambda} = \displaystyle\frac{\lambda - \lambda_c(N,m,k)}{\sigma_\lambda(N,m,k)} \;.
\label{eq-lam-sc}
\ee 
We choose $N = 5$ and $m = 10$ system as the distributions may have smooth form only in the asymptotic limit with $m >> N$ for bosons.  Note that the modified Gumbel (smooth red curves) distributions are obtained using Eq. \eqref{gumbel}. Similarly, TW (smooth blue curves) distributions are obtained using Eq. \eqref{eq-tw}.  It is important to mention that unlike 1000 member ensemble used in Section \ref{sec2},  here 5000 member ensemble is used so that the predictions are robust for the change,  as $k$ (or $q$ varies) varies,  in the lowest eigenvalue distribution.

\begin{figure}[tbh]
	\begin{center}
		\includegraphics[width=0.5\linewidth]{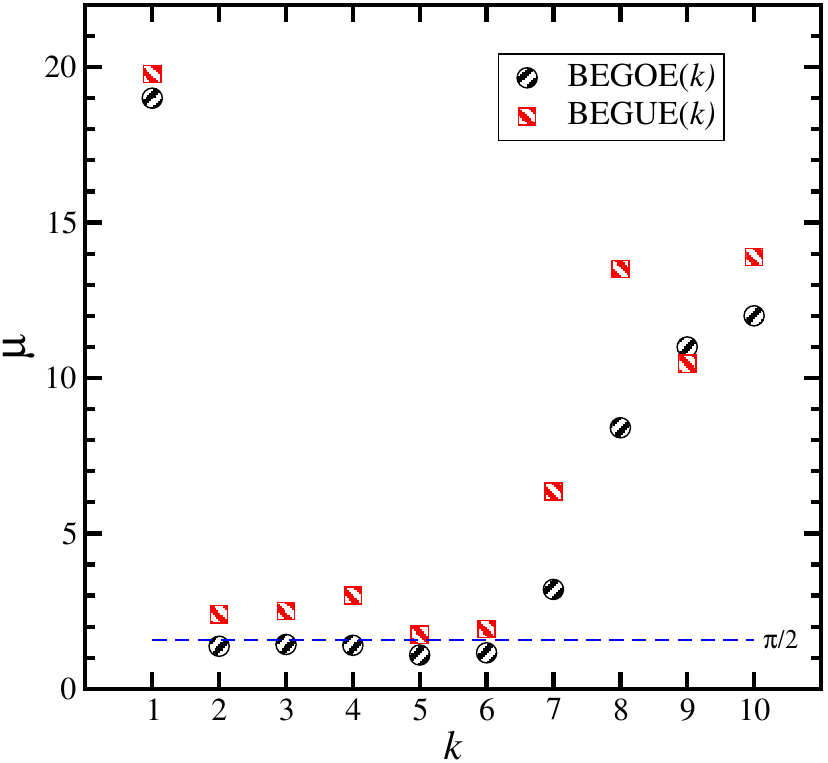}
	\end{center}
	\caption{Values of the fitting parameter $\mu$ for modified Gumbel ($G_\mu$) distribution fits for BEGOE($k$) and BEGUE($k$). The horizontal dashed line at $\mu = \pi/2$ corresponds to modified Gumbel distribution applicable to different correlated systems, as conjectured in \cite{Bram-00}.}
		\label{fig-new2}
\end{figure}

\begin{figure}[H]
	\begin{center}
		\includegraphics[width=0.5\linewidth]{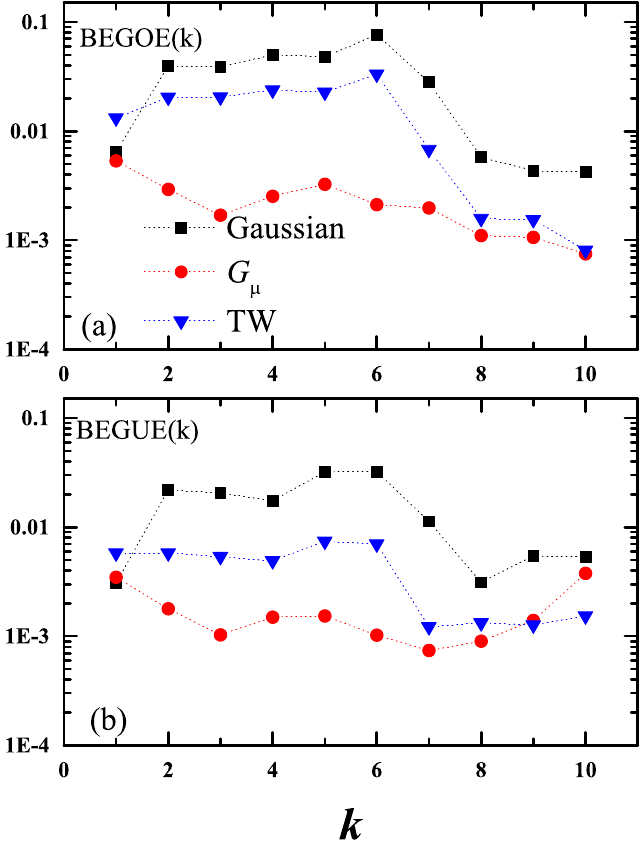}
	\end{center}
	\caption{Residual sum of squares for Gaussian,  modified Gumbel ($G_\mu$) and TW distribution fits shown in Figs. \ref{fig5} and \ref{fig6} for (a) BEGOE($k$) and (b) BEGUE($k$).}
		\label{fig-new}
\end{figure}

For $k=1$, the distributions are close to Gaussian form. However, for $2 \leq k \leq 6$ the distributions are close to modified Gumbel form and it transitions to TW for $k = 10$.  The values of the fitting parameter $\mu$ are shown in Fig.  \ref{fig-new2}.  Note that $\mu$ values for $k =$ $2, \;3,\ldots,6$ are very close to the value conjectured for correlated systems \cite{Bram-00}.  We also analyzed residual sum of squares shown in Fig. \ref{fig-new}, which also confirmed that modified Gumbel is the best fit curve for $k=2-6$. Thus, the lowest eigenvalue distribution for BEGOE($k$)/BEGUE($k$) changes from Gaussian to modified Gumbel to TW as $q(N,m,k)$ changes from $\sim 1$ ($k = 1$) to 0 ($k = m$).  For fermion systems, it was concluded that there is a transition from Gaussian to TW form for the largest eigenvalue distribution \cite{Be-23}.

\begin{figure}[tbh]
	\begin{center}
		\includegraphics[width=\linewidth]{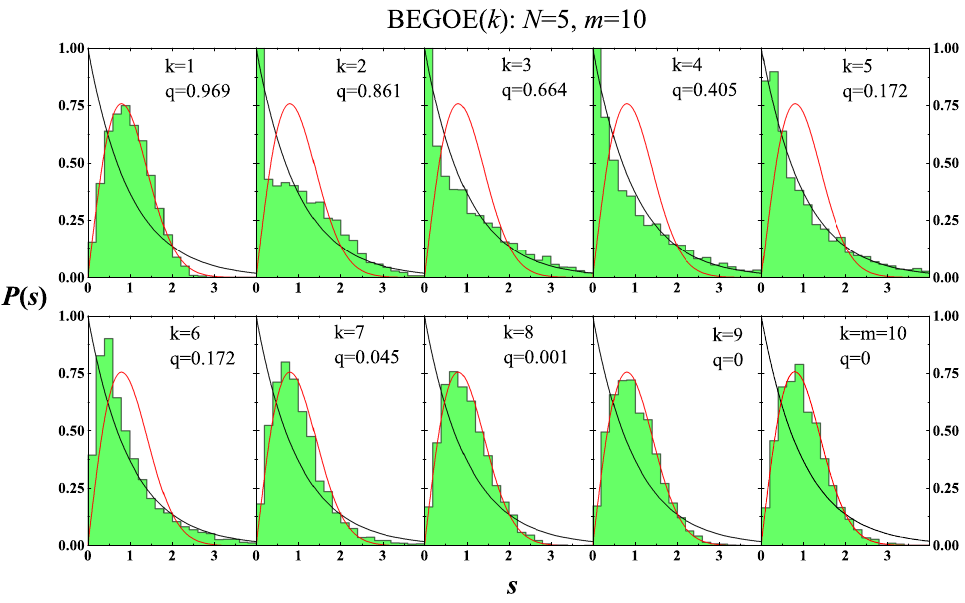}
	\end{center}
	\caption{Distribution of normalized spacings $s$ between the lowest and the next lowest eigenvalue for the same system considered in Fig. \ref{fig5}. We compare the numerical histograms with Poisson distribution (black smooth curve) and the Wigner's surmise (red smooth curve) for GOE \cite{Mehta}.}
		\label{fig7}
\end{figure}

\begin{figure}[tbh]
	\begin{center}
		\includegraphics[width=\linewidth]{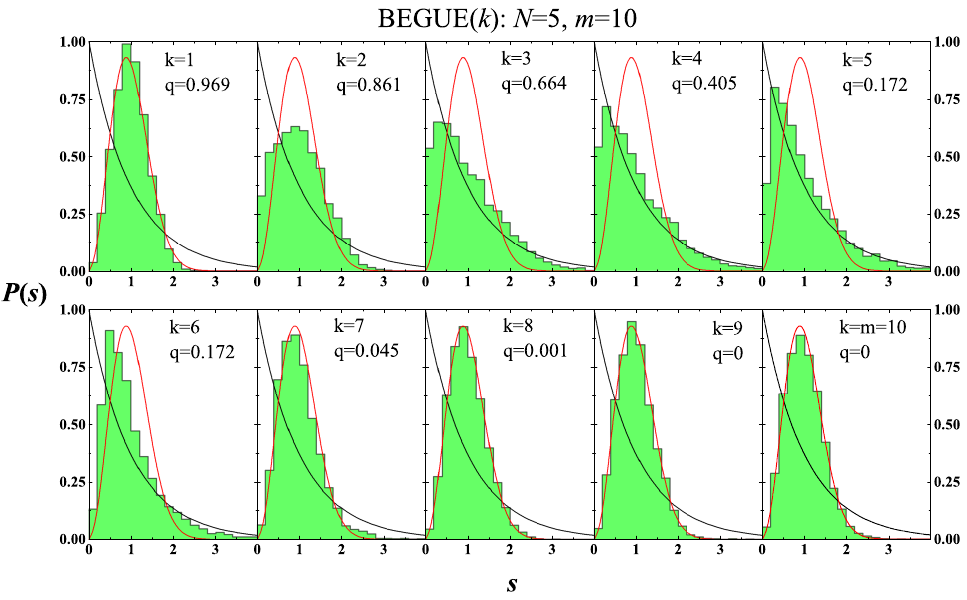}
	\end{center}
	\caption{Same as Fig.~\ref{fig7} but for BEGUE($k$). Note that the corresponding Wigner's surmise is for GUE \cite{Mehta}.}
	\label{fig8}
\end{figure}

We also studied the distribution of the spacings between lowest and the next lowest eigenvalue, which is also an extreme statistic.  We use the normalized spacing $s$ which is the ratio of actual spacing with the average spacing and the results are shown in Figs. \ref{fig7} and \ref{fig8} respectively for BEGOE($k$) and BEGUE($k$).  Here, the average spacing is calculated over all the members of the ensemble. The numerical histograms are compared with Poisson and the Wigner's surmise results. It shows that as $q$ parameter changes, there is a transition from Wigner's surmise (for $k = 1$) to Poisson (for $k = 2-4$) to Wigner's surmise (starting from $k = 5$ to $k =10$).  This indicates that lowest and the next lowest eigenvalue are statistically independent for large $s$ for $k=1$ and for $k=5$ to $k =10$ \cite{Perret14}.  While for $k = 2-4$, the departure from GOE/GUE is expected as shown in \cite{SeligmanFlores}.  The spacings at the spectrum edge thus behave differently from the spacings inside the spectrum bulk.  Analytical understanding of this \cite{PRL-23, Sa-24} is also an essential ingredient for derivation of EVS for EE. 

\section{Conclusions and future outlook}
\label{sec4}

We have presented numerical results for the first four moments of the lowest eigenvalue distribution for BEGOE($k$) and BEGUE($k$). We have analytical understanding of the centroid,  from the $q$-normal form of the ensemble averaged eigenvalue density for EE($k$),  and for the other three moments, one needs to derive analytical results.  We have also tested ansatz for variances based on the $q$-normal form of the ensemble averaged eigenvalue densities and the result for the variance of TW form.  These ansatz are found to give good estimates.  In order to be able to derive analytical results,  one first has to derive the distribution of correlated matrix elements for EE as a function of parameter $q$ (or equivalently, interaction rank $k$).  Qualitatively,  larger(smaller) values of the parameter $q$ implies pre-dominance of few-body(many-body) interactions in the complex system under consideration.  As a first step in gaining insight into the transition in distribution of smallest eigenvalue with decreasing value of parameter $q$,  we employ numerical approach.  

Numerical results suggest that the distribution of the lowest eigenvalues for BEGOE($k$) and BEGUE($k$) make a transition from Gaussian $[q(N,m,k) \sim 1]$ to modified Gumbel (intermediate $q(N,m,k)$ values) to TW $[q(N,m,k) = 0]$ as we change the rank of interactions $k$.  We have confirmed this by analyzing the residual sum of square values.  The values of the fitting parameter $\mu$ defining modified Gumbel distributions for $k =$ $2, \;3,\ldots,6$ are very close to the value conjectured for correlated systems \cite{Bram-00}.  Although not shown in Figs.  \ref{fig5} and \ref{fig6},  we also used $\chi^2$ distributions \cite{Ch-14,  PRE-22} to analyze this transition and it gives a good fit for all $q$ values.  As we do not have good understanding of the physical significance of $\chi^2$ form for EVS for EE, we have dropped the $\chi^2$ distributions from the discussion. 

Similarly, the distribution of normalized spacing between the lowest and next lowest eigenvalues exhibits a transition from Wigner's surmise (for $k = 1$) to Poisson (for $k = 2-4$) to Wigner's surmise (starting from $k = 5$ to $k =10$) with decreasing $q$ value.  The spacings at the spectrum edge thus behave differently from the spacings inside the spectrum bulk.  Analytical understanding of this \cite{PRL-23, Sa-24} is also an essential ingredient for derivation of EVS for EE.  It should be emphasized that the analysis presented in this paper and its  connection with the $q$-normal form of eigenvalue densities is novel.  Our results also show a strong dependency of moments of the smallest eigenvalue densities on ergodicity,  as expected,  which has not been demonstrated earlier; see Fig.  \ref{fig9} and related discussion for details.  The set of numerical calculations presented in this paper for BEGOE($k$) and BEGUE($k$) may be used as a starting point for further exploring the EVS in random matrix ensembles appropriate for quantum many-body interacting systems.  In conclusion,  our results for EVS for bosonic EE bring out the role of the so-called $q$ parameter that is central to $k$-body EE ($q$ changes from 1 to 0 as $k$ changes from 1 to $m$).

\section{Acknowledgments}

M. V.  acknowledges financial support from CONAHCYT project Fronteras 10872.

\ed
\end{document}